\documentstyle[sprocl]{article}

\def\be{\begin{equation}}
\def\ee{\end{equation}}
\def\bea{\begin{eqnarray}}
\def\eea{\end{eqnarray}}
\newcommand{\OO}{{\cal O}}
\newcommand{\sss}[1]{\mbox{\scriptsize #1}}
\newcommand{\M}{{\cal M}}

\newcommand{\PP}{{\cal P}}
\newcommand{\D}{{\cal D}}

\newcommand{\GeV}{\unskip\,\mathrm{GeV}}
\newcommand{\MeV}{\unskip\,\mathrm{MeV}}

\begin{document}

\title{RADIATIVE CORRECTIONS TO W-PAIR MEDIATED FOUR-FERMION PRODUCTION AT 
       LEP2}

\author{W.~BEENAKKER}

\address{Physics Department, University of Durham, Durham DH1 3LE, England} 

\author{F.A.~BERENDS\ \ \ and\ \ \ A.P.~CHAPOVSKY}

\address{Instituut--Lorentz, University of Leiden, P.O.~Box 9506, 
         2300 RA Leiden,\\
         The Netherlands}


\maketitle\abstracts{We present the radiative corrections to off-shell
                     $W$-pair production, as calculated within the double-pole 
                     approximation.}

\section{Introduction}

This report summarizes the results on electroweak radiative corrections to 
reactions that involve the production and subsequent decay of pairs of $W$
bosons, which are of prime interest at LEP2. Unless stated otherwise we use 
the reaction
\be
\label{process}
  e^{+}(q_{1})\,e^{-}(q_{2})
  \to
  \mu^{+}(k_{1})\,\nu_{\mu}(k_{1}')\,\tau^{-}(k_{2})\,\bar{\nu}_{\tau}(k_{2}')
\ee
as typical example, but all four-fermion final states originating from
$W$-boson pairs can be treated in a similar way. Since the RADCOR~98 
conference took place our results have been finalized and discussed 
elsewhere~\cite{ww-dpa}. A highlight of the main points is given here.
For related issues, see these proceedings~\cite{dittm-proc}.

A complete calculation of $\OO(\alpha)$ radiative corrections to a process 
with six external particles is beyond present possibilities. However, the 
possible presence of two resonances as intermediate state in reaction 
(\ref{process}) offers an additional expansion parameter $\Gamma_{W}/M_{W}$ 
besides $\alpha$. Since the ratio between the width and mass of the
$W$ boson  is also of order $\alpha$, a double expansion both in $\alpha$ and 
$\Gamma_{W}/M_{W}$ is a natural and economic way to simplify the calculation. 
Thus the aim is to calculate radiative corrections of $\OO(\alpha)$ and  
$\OO(\Gamma_{W}/M_{W})$, while neglecting higher-order terms such as 
$\alpha \Gamma_{W}/M_{W}$. The way in which the expansion is carried out is 
called the double-pole approximation~\cite{pole-scheme} (DPA), which possesses 
the important feature of maintaining gauge invariance.

\section{The Born cross-section in the double-pole approximation}

As an illustration, we shall first apply the DPA method to the Born 
cross-section. When one calculates the Born cross-section $d\sigma^{0}$ in 
the usual way, gauge invariance demands not only the inclusion of all diagrams 
but also a gauge-invariant procedure to incorporate the width in the $W$-boson 
propagators. For a discussion of these issues we refer to the 
literature~\cite{lep2,ww-born}. Numerically the three diagrams involving two 
resonant $W$-boson propagators (called CC03) are the dominant ones. The 
diagrams containing a single resonant $W$ boson or no resonant $W$ boson at 
all are suppressed by $\Gamma_{W}/M_{W}$ and $(\Gamma_{W}/M_{W})^{2}$,
respectively, although the exchange of almost real photons may enhance
suppressed terms. The CC03 part of the cross-section is not gauge-invariant, 
but its residue in the DPA is. Therefore we split up the lowest-order 
cross-section according to
\be
\label{born+bkg}
  d\sigma^{0}
  =
  d\sigma^{0}_{\sss{DPA}} 
  +
  \bigl( d\sigma^{0} - d\sigma^{0}_{\sss{DPA}} \bigr),
\ee
where the first term originates from the CC03 amplitude in the DPA limit and
the second term is suppressed by at least $\Gamma_{W}/M_{W}$.

The way in which $d\sigma_{\sss{DPA}}^{0}$ is defined starts with the CC03 
amplitude
\be
\label{pole/A}
  \M 
  =
  \sum_{\lambda_{1},\lambda_{2}}
  \Pi_{\lambda_{1}  \lambda_{2}}(M_{1}, M_{2}) \
  \frac{\Delta^{(+)}_{\lambda_{1}}(M_{1})}{D_{1}} \
  \frac{\Delta^{(-)}_{\lambda_{2}}(M_{2})}{D_{2}} \ ,
\ee
where the summation runs over the $W^{\pm}$ helicities ($\lambda_{1,2}$).
The quantities $\Delta^{(\pm)}$ and $\Pi$ are the off-shell amplitudes
for $W^{\pm}$ decay and $W$-pair production, respectively.
The inverse propagators $D_i$ ($i=1,2$) are defined as
\be
  D_{i}=M_{i}^{2}-M_{W}^{2}+i\Gamma_{W}M_{W},
  \quad\quad 
  M_{i}^{2}=(k_{i}+k_{i}')^{2}.
\ee
The momenta of the resonant $W^{\pm}$ bosons will be indicated by $p_{1,2}$.
Introducing the production/decay phase-space factors
\begin{eqnarray}
  d\Gamma_{\sss{pr}}\
  &=& \frac{1}{(2\pi)^{2}}\, \delta(q_{1}+q_{2}-p_{1}-p_{2})\,
      \frac{d \vec{p}_{1}}{2p_{10}}\,
      \frac{d \vec{p}_{2}}{2p_{20}}, 
      \\
  d\Gamma_{\sss{dec}}^{+}
  &=& \frac{1}{(2\pi)^{2}}\, \delta(p_{1}-k_{1}-k_{1}')\,
      \frac{d \vec{k}_{1}}{2k_{10}}\,
      \frac{d \vec{k}_{1}'}{2k_{10}'}, 
\end{eqnarray}
and a similar expression for $d\Gamma_{\sss{dec}}^{-}$, the differential  
cross-section reads 
\begin{eqnarray}
\label{pole/born}
  d\sigma
  &=& \frac{1}{2s}\sum\limits_{\lambda_1,\lambda_2,\lambda_1',\lambda_2'}
      \PP_{[\lambda_{1}\lambda_{2}][\lambda_{1}'\lambda_{2}']}(M_{1},M_{2})\,
      d\Gamma_{\sss{pr}}
      \times
      \D_{\lambda_{1}\lambda_{1}'}(M_{1})\, d\Gamma_{\sss{dec}}^{+}
      \times \nonumber \\
  & & \times
      \D_{\lambda_{2}\lambda_{2}'}(M_{2})\, d\Gamma_{\sss{dec}}^{-}
      \times\, 
      \frac{1}{2\pi}\,\frac{d M_{1}^{2}}{|D_{1}|^{2}}
      \times
      \frac{1}{2\pi}\,\frac{d M_{2}^{2}}{|D_{2}|^{2}},
\end{eqnarray}
where
\be
\label{pole/prod}
  \PP_{[\lambda_{1}\lambda_{2}][\lambda_{1}'\lambda_{2}']}(M_{1},M_{2})\ \ 
  =
  \sum\limits_{\sss{$e^{\pm}$ helicities}}
  \Pi_{\lambda_{1}\lambda_{2}}(M_{1},M_{2}) \
  \Pi_{\lambda_{1}'\lambda_{2}'}^{*}(M_{1},M_{2}),
\ee
\be
\label{pole/decay}
  \D_{\lambda_{i}\lambda_{i}'}(M_{i})\ \ 
  =
  \sum\limits_{\sss{fermion helicities}}
  \Delta_{\lambda_{i}}(M_{i}) \
  \Delta_{\lambda_{i}'}^{\!*}(M_{i}).
\ee
For the double-pole approximation of Eq.~(\ref{pole/born}) we choose the 
prescription $p_{i}^{2}=M_{i}^{2}\to M_{W}^{2}$ both in the phase-space 
factors and in $\PP$ and $\D$, which can be identified as on-shell density 
matrices. Strictly speaking, the double-pole residue should be taken at the
complex pole, $M_{i}^{2}\to M_{W}^{2}-i\Gamma_{W}M_{W}$. In order to avoid 
problems with complex kinematics and complicated analytic continuations,
however, we use the limit $M_{i}^{2}\to M_{W}^{2}$ instead. In our procedure 
the phase-space factors are given by the on-shell expressions in the 
laboratory system, with the production/decay angles being free and the
energies being fixed by the DPA limit. This is achieved by writing $d\sigma$ 
as a differential in both $M_{i}^{2}$ and the solid production/decay angles. 
Distributions in $M_{i}^{2}$ are solely determined by 
$d M_{i}^{2}/|D_{i}|^{2}$ and not anymore by the matrix element, at least not
at Born level. The full integration over the $M_{i}^{2}$ distribution is 
approximated by 
\be
  \label{mass_integration}
  \frac{1}{2\pi} \int\limits_{-\infty}^{\infty} d M_{i}^{2} 
  \frac{1}{|D_{i}|^{2}}
  =
  \frac{1}{2M_{W}\Gamma_{W}}.
\ee
For distributions in other variables, like energies, an appropriate off-shell
Jacobian should be inserted.

In the following we shall consider the radiative corrections to 
$d\sigma^{0}_{\sss{DPA}}$, so that the $\OO(\alpha)$ and $\OO(\Gamma_W/M_W)$ 
corrected cross-section is obtained as 
\be
\label{born+dpa}
  d\sigma
  =
  d\sigma^{0}_{\sss{DPA}} (1 + \delta_{\sss{DPA}})
  +
  \bigl( d\sigma^{0} - d\sigma^{0}_{\sss{DPA}} \bigr).
\ee

\section{Radiative corrections in the double-pole approximation}

Besides the usual distinction between virtual and real corrections one should 
now also distinguish between factorizable and non-factorizable corrections. 
The factorizable corrections are corrections that apply to the production or 
decay parts separately, whereas the non-factorizable corrections can be 
regarded as interference effects between different stages of the off-shell 
process.

The factorizable corrections comprise the on-shell corrections to the $W$-pair
production process and the on-shell corrections to the partial decay widths of
the $W$ bosons, generalized to the density matrices $\PP$ and $\D$. We have
calculated these generalized corrections by extending existing calculations 
for $W$-pair production~\cite{ww-prod} and decay~\cite{w-dec}. For virtual 
corrections and soft-photon radiation [$E_{\gamma} \ll \Gamma_W$] no further 
clarifications are required.

The radiation of more energetic photons, however, requires some special care 
in the treatment of both the matrix element and the phase space. When a photon
with momentum $k$ is emitted from one of the intermediate $W$ bosons, the
resulting product of $W$-boson propagators can be written as 
\be
\label{decomp}
  \frac{1}{D_i D_{i\gamma}}
  =
  \frac{1}{2kp_i} \Biggl[\frac{1}{D_i}-\frac{1}{D_{\i\gamma}}\Biggr],
\ee
where
$$
  D_{i(i\gamma)}=M_{i(i\gamma)}^2-M_W^2+i\Gamma_{W}M_{W},
$$
\be
  M_i^2=(k_i+k_i')^2,
  \quad\quad
  M_{i\gamma}^2=(k_i+k_i'+k)^{2},
  \quad\quad   
  p_i=k_i+k_i'(+k).
\ee
The first term on the r.h.s of Eq.~(\ref{decomp}) can now be interpreted as 
belonging to the production of a $W$ boson plus a photon, whereas the second 
term corresponds to the radiative decay of the $W$ boson. Therefore, the 
real-photon bremsstrahlung cross-section has the following structure
\be
\label{rad/mx2}
  d\sigma
  =
  \frac{1}{2s}\,|\M_{\gamma}|^{2}\,d\Gamma_{4f\gamma} 
  =
  \frac{1}{2s}\,|\M_{0}+\M_{+}+\M_{-}|^{2}\,d\Gamma_{4f\gamma}, 
\ee
with $d\Gamma_{4f\gamma}$ indicating the complete off-shell phase-space factor.
The exact amplitude in Eq.~(\ref{rad/mx2}) has been divided into parts related 
to radiative $W$-pair production ($\M_{0}$) and radiative $W$-boson decays
($\M_{\pm}$). 

For hard photons [$E_{\gamma} \gg \Gamma_{W}$] the interference terms between 
these different radiation sources are suppressed, since the propagators peak 
in different parts of phase space. Thus merely the quadratic 
$|\M_{0,\pm}|^{2}$ parts need to be retained in Eq.~(\ref{rad/mx2}).
Introducing the radiative production/decay phase-space factors 
\be
  d\Gamma_{\sss{pr}}^{\gamma}
  =
  \frac{1}{(2\pi)^{2}}\,
  \delta(q_{1}+q_{2}-p_{1}-p_{2}-k)\, \frac{d\vec{p}_{1}}{2p_{10}}\, 
  \frac{d\vec{p}_{2}}{2p_{20}}\,    \frac{d\vec{k}}{(2\pi)^{3}2k_{0}},
\ee
\be
  d\Gamma_{\sss{dec}}^{+\gamma}
  =
  \frac{1}{(2\pi)^{2}}\,
  \delta(p_{1}-k_{1}-k_{1}'-k)\, \frac{d\vec{k}_{1}}{2k_{10}}\, 
  \frac{d\vec{k}_{1}'}{2k_{10}'}\,  \frac{d\vec{k}}{(2\pi)^{3}2k_{0}},
\ee
and a similar expression for $d\Gamma_{\sss{dec}}^{-\gamma}$, the complete 
off-shell phase-space factor $d\Gamma_{4f\gamma}$ can be written in three
equivalent forms
\be
  d\Gamma_{4f\gamma} 
  = 
  d\Gamma_{\sss{0}}^{\gamma} 
  = 
  d\Gamma_{\sss{pr}}^{\gamma} \cdot 
  d\Gamma_{\sss{dec}}^{+} \cdot d\Gamma_{\sss{dec}}^{-} \cdot   
  \frac{dM_{1}^{2}}{2\pi} \cdot \frac{d M_{2}^{2}}{2\pi},
\ee
\be
  d\Gamma_{4f\gamma} 
  = 
  d\Gamma_{+}^{\gamma} 
  = 
  d\Gamma_{\sss{pr}} \cdot 
  d\Gamma_{\sss{dec}}^{+\gamma} \cdot d\Gamma_{\sss{dec}}^{-} \cdot   
  \frac{d M_{1\gamma}^{2}}{2\pi} \cdot \frac{d M_{2}^{2}}{2\pi},
\ee
and a similar expression for $d\Gamma_{-}^{\gamma}$. Consequently the
hard-photon cross-section can be rewritten as
\be
  d\sigma
  =
  \frac{1}{2s}
  \biggl[  |\M_{0}|^{2}\,d\Gamma_{0}^{\gamma}
          +|\M_{+}|^{2}\,d\Gamma_{+}^{\gamma}
          +|\M_{-}|^{2}\,d\Gamma_{-}^{\gamma}
  \biggr].
\ee
When the appropriate DPA limits are taken, three distinct factorizable
contributions emerge, with the photon assigned to either the production stage
or one of the decay stages. 

For semi-soft photons [$E_{\gamma} = \OO(\Gamma_{W})$] the matrix element can 
be written in a factorized soft-photon form with the propagators $D_{i}$ and 
$D_{i\gamma}$ kept exact. This has two consequences. One is that in this regime
$|\M_{\pm}|^{2}$ can be replaced by simplified expressions, which are useful 
for discussing final-state radiation (FSR) effects on the resonance 
shapes~\cite{ww-dpa,fsr}. The second consequence concerns the non-factorizable
interference terms in Eq.~(\ref{rad/mx2}) between the different radiative
production and decay stages. These interference terms cannot be neglected
anymore in the semi-soft regime, as the propagators $D_{i}$ and $D_{i\gamma}$ 
start to overlap. In a similar way, also the virtual non-factorizable 
corrections originate solely from semi-soft photonic interconnection effects 
between the different production and decay stages. For detailed discussions of 
the non-factorizable corrections we refer to the 
literature~\cite{nf-my,nf-bbc,nf-ddr}. We merely state here that their
numerical impact is in general relatively small.

\section{Numerical results}

For the numerical evaluations we adopt the LEP2 input-parameter
scheme~\cite{lep2}. The independent input parameters are the Fermi constant 
$G_{\mu}$, the fine-structure constant $\alpha$, the masses of the light 
fermions (which reproduce the experimentally determined hadronic vacuum 
polarization), and the masses of the $W,Z$ bosons. Subsequently, we make use of
the Standard Model prediction for $G_{\mu}$ in terms of the above parameters, 
the top-quark mass $m_{t}$, the Higgs-boson mass $M_{H}$, and the strong 
coupling $\alpha_{\sss{S}}$. In this way $m_{t}$ will be fixed once the other
parameters have been chosen. In the lowest-order cross-sections we use 
$G_{\mu}$ instead of $\alpha$ and compensate for this replacement in the 
one-loop corrections, i.e.~we use the so-called $G_{\mu}$ parametrization.
 
In the following we present a small selection of interesting numerical results 
as obtained with our DPA procedure~\cite{ww-dpa}.

\begin{figure}
  \unitlength 1cm
  \begin{center}
  \begin{picture}(4,5)
  \put(-2,4){\makebox[0pt][c]{\boldmath $\frac{\displaystyle d\sigma}
                                 {\displaystyle dE_{\gamma}}$}}
  \put(-2,3){\makebox[0pt][c]{\bf [fb]}}
  \put(6,0.5){\makebox[0pt][c]{\bf\boldmath $E_{\gamma}$ [GeV]}}
  \put(-2.5,-3){\includegraphics{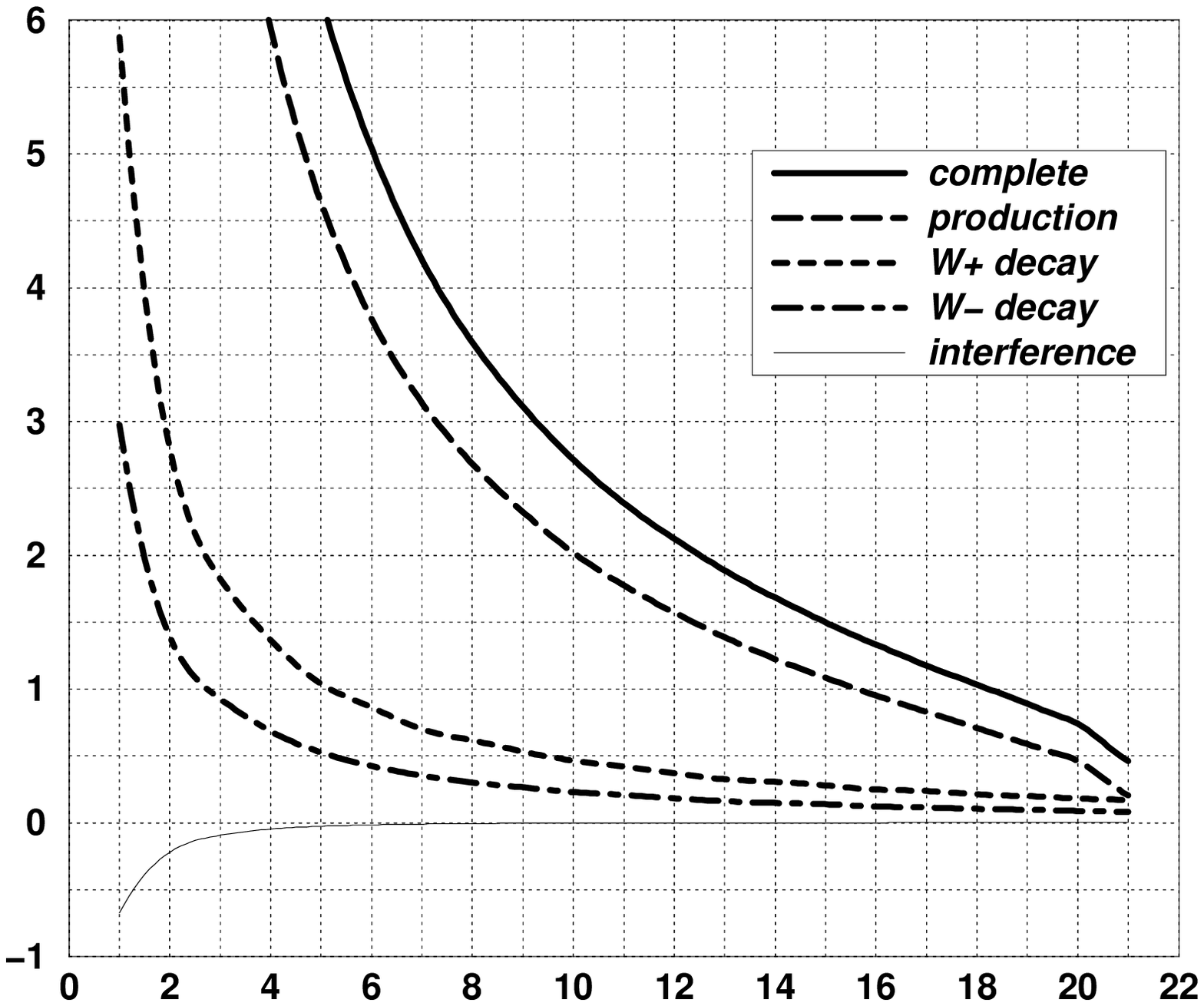}}
  \end{picture}
  \end{center}
  \caption[]{The photon-energy distribution $d\sigma/dE_{\gamma}$ for the
             $\mu^+\nu_{\mu} \tau^-\bar{\nu}_{\tau} \gamma$ final state at 
             the LEP2 energy of $184\GeV$. In addition the separate production 
             and decay contributions are given.}
\label{fig:rad}
\end{figure}%
In Fig.~\ref{fig:rad} we start off with the DPA energy spectrum of the photon 
($d\sigma/dE_{\gamma}$) and the separate contributions to it from the 
production stage, the decay stages, and the semi-soft interference terms. 
As can be seen clearly, the contribution from the $W^{+}$ decay comes out 
larger than the one from the $W^{-}$ decay. This is a result of the fact that 
we consider reaction~(\ref{process}) with its specific choice of charged 
leptons. It simply means that the photon-energy spectrum originating from the 
decay $W \to \mu\nu_{\mu}\gamma$ exceeds the one from 
$W \to \tau\nu_{\tau}\gamma$, as expected from collinear-photon considerations.

For inclusive photons, the corrections to the total cross-section and the 
($W^{\pm}$) production-angle distribution are essentially the same as for 
stable $W$ bosons, provided one treats the widths in the propagators and in
the decay channels in the same way. As is well known, these corrections are 
dominated by the large logarithms of initial-state radiation (ISR). If one 
does not integrate over the invariant masses $M_{1}$ and/or $M_{2}$, one will
in addition observe FSR and non-factorizable effects. The latter
effects are in general relatively small (see e.g.~ Fig.~\ref{fig:rad}).
The FSR effects, however, may lead to a sizeable distortion of the
invariant-mass distributions~\cite{fsr}. 

\begin{figure}
  \unitlength 1cm
  \begin{center}
  \begin{picture}(4,5)
  \put(-2.5,4){\makebox[0pt][c]{\boldmath $\frac{\displaystyle d\sigma}
                                 {\displaystyle dM_1^2}$}}
  \put(-2.5,3){\makebox[0pt][c]{\bf [fb]}}
  \put(6.5,0.5){\makebox[0pt][c]{\bf\boldmath $M_1$ [GeV]}}
  \put(-2.5,-3){\includegraphics{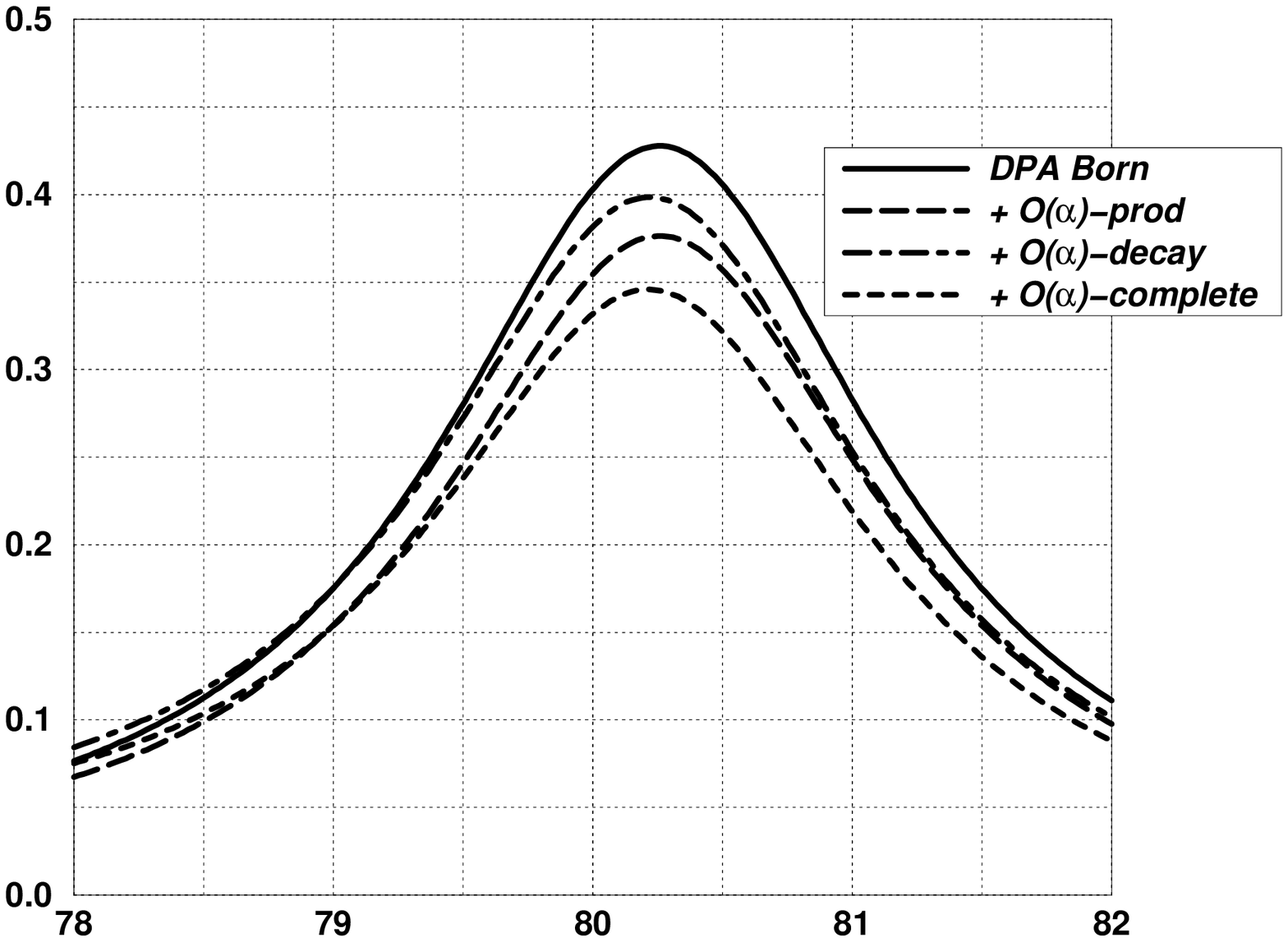}}
  \end{picture}
  \end{center}
  \caption[]{The invariant-mass distribution $d\sigma/dM_1^2$ for the
             $\mu^+\nu_{\mu} \tau^-\bar{\nu}_{\tau}$ final state
             at the LEP2 energy of $184\GeV$.} 
\label{fig:mass}
\end{figure}%
\begin{figure}
  \unitlength 1cm
  \begin{center}
  \begin{picture}(4,5)
  \put(-2,4){\makebox[0pt][c]{\bf\boldmath $\delta_{\sss{DPA}}$}}
  \put(-2,3){\makebox[0pt][c]{\bf [\%]}}
  \put(6.5,0.5){\makebox[0pt][c]{\bf\boldmath $M_1$ [GeV]}}
  \put(-2.5,-3){\includegraphics{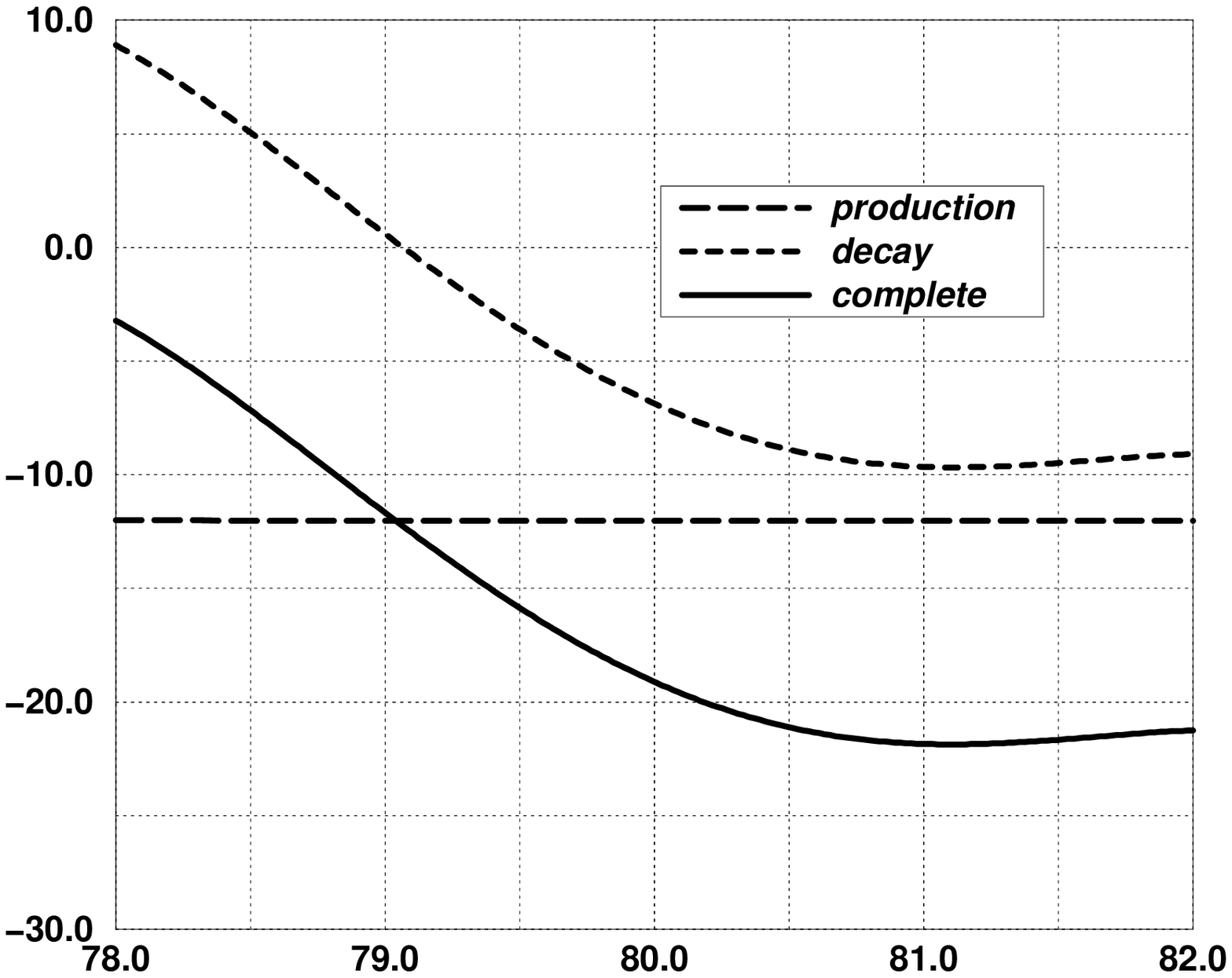}}
  \end{picture}
  \end{center}
  \caption[]{The relative correction factors $\delta_{\sss{DPA}}$ 
             corresponding to Fig.~\protect\ref{fig:mass}.}
\label{fig:mass/rel}
\end{figure}%
The FSR distortion effects are displayed in Figs.~\ref{fig:mass} and 
\ref{fig:mass/rel} for the invariant-mass distribution $d\sigma/dM_1^2$ of 
reaction~(\ref{process}), which involves the decay $W^+ \to \mu^{+}\nu_{\mu}$. 
The effect decreases for the decay $W^+ \to \tau^{+}\nu_{\tau}$ and increases 
for $W^+ \to e^{+}\nu_{e}$. The corresponding shift in the peak position of 
the Breit--Wigner line shape is found to be $-20,-39,$ and $-77\MeV$ for the
decays $W^+ \to \tau^{+}\nu_{\tau}$, $\mu^{+}\nu_{\mu}$, and $e^{+}\nu_{e}$,
respectively. These large distortion effects hinge on the fact that one is 
able to determine $M_{1}$ for a dilepton final state. In practical experimental
situations the effect wil be smaller, for instance because a photon emitted
collinear with the charged lepton may not be separately detectable.
Nevertheless it is useful to be aware of this effect.
\newpage

\begin{figure}
  \unitlength 1cm
  \begin{center}
  \begin{picture}(4,5)
  \put(-4,2.5){\makebox[0pt][c]{\bf\boldmath $\delta_{\sss{DPA}}$ [\%]}}
  \put(6.6,1){\makebox[0pt][c]{\boldmath $M_{2}$ $[\GeV]$}}
  \put(0,0.3){\makebox[0pt][c]{\boldmath $M_{1}$ $[\GeV]$}}
  \put(-3,-1){\includegraphics{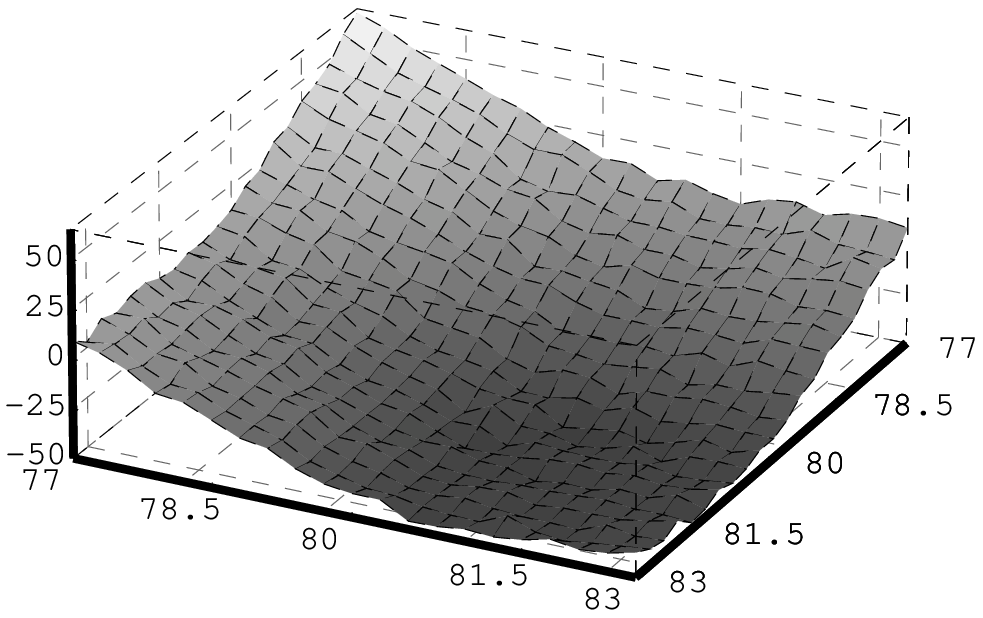}}
  \end{picture}
  \end{center}
  \caption[]{Correction to the double invariant-mass distribution
             $\,d\sigma/dM_1^2 dM_2^2\,$ for the $e^+\nu_{e} e^-\bar{\nu}_{e}$
             final state at the LEP2 energy of $184\GeV$.}
\label{fig:mm/rel/ee}
\end{figure}%
\begin{table}
\[
  \begin{array}{|c|c c c|}    \hline 
    \multicolumn{4}{|c|}{\raisebox{-1mm}{$\delta^+_{\sss{dec}}(M_1)$}}\\[2mm] 
    \hline 
    \raisebox{-3mm}{$\Delta_1$} 
    & \multicolumn{3}{|c|}{\raisebox{-1mm}{decay channel}}\\ 
    \cline{2-4}
         & \ \ e^{+}\nu_e\ \  & \ \ \mu^{+}\nu_{\mu} \ \  
         & \ \ \tau^{+}\nu_{\tau} \ \   \\ 
    \hline 
    \ \ -1/2 \ \ & -1.4       & -0.8              & -0.5  \\
     0   & -15.0      & -7.8              & -4.0  \\
     1/2 & -17.3      & -9.0              & -4.6  \\
    \hline
  \end{array} 
  \quad\quad 
   \begin{array}{|c|c c c|}    \hline 
    \multicolumn{4}{|c|}{\raisebox{-1mm}{$\delta_{\sss{nf}}(M_1,M_2)$}}\\[2mm] 
    \hline  
    \raisebox{-3mm}{$\Delta_1$} 
    & \multicolumn{3}{|c|}{\raisebox{-1mm}{$\Delta_2$}}\\ 
    \cline{2-4}
         & \ \ -1/2 \ \  & \ \  0  \ \   & \ \ 1/2 \ \   \\ 
    \hline 
    \ \ -1/2 \ \ & +0.5  & +0.2  & -0.1  \\
     0   & +0.2  & +0.0  & -0.2  \\
     1/2 & -0.1  & -0.2  & -0.4  \\
    \hline 
  \end{array}
\]
\caption[]{Relative correction factors [in \%] for the double invariant-mass 
           distribution $\,d\sigma/dM_1^2 dM_2^2\,$ at the LEP2 energy of
           $184\GeV$. 
           Left: the corrections from the $W^+$-boson decay stage 
           $\delta_{dec}^{+}(M_1)$ for different leptonic decay channels. 
           Right: the non-factorizable corrections 
           $\delta_{\sss{nf}}(M_1,M_2)$. 
           Three near-resonant invariant masses are considered: 
           $\Delta_i = (M_i-M_W)/\Gamma_W = -1/2,0,1/2$.}
\label{tab:3}
\end{table}%
For a double invariant-mass distribution $\,d\sigma/dM_1^2 dM_2^2\,$ the 
distortion becomes even more pronounced, as can be seen from 
Fig.~\ref{fig:mm/rel/ee} where the $e^+\nu_{e} e^-\bar{\nu}_{e}$ final state
is considered. Explicit numbers can be found in Table~\ref{tab:3}, where
we have split up the relative correction factor into production, decay, and 
non-factorizable contributions according to 
$$
  \delta_{\sss DPA}(M_1,M_2)
  =
  \delta_{\sss{pr}} + \delta_{\sss{dec}}^{+}(M_1)
  + \delta_{\sss{dec}}^{-}(M_2) + \delta_{\sss{nf}}(M_1,M_2).
$$
Here the correction $\delta_{\sss{pr}}$ from the production part is independent
of $M_i$ and equals $-12.0\%$ at the considered energy of $184\GeV$.

\section{Conclusions}

We have summarized the gauge-invariant DPA calculation of the $\OO(\alpha)$ 
and $\OO(\Gamma_{W}/M_{W})$ radiative corrections to $W$-pair mediated 
four-fermion production. Special attention has been paid to a practical 
implementation of the DPA scheme and to the definition of the so-called 
factorizable and non-factorizable corrections. This DPA procedure can also
be used for other reactions where unstable particles are produced in pairs.

\section*{Acknowledgements}

We acknowledge the support by PPARC (WB) and by the Stichting FOM (APC).
FB would like to thank the organizers for their kind invitation to a 
stimulating conference. Moreover, we are indebted to Dr.~S.~Dittmaier for 
helpful discussions and the exchange of results.

\section*{References}

\newcommand{\np}[3]{{\sl Nucl. Phys.} {\bf #1} (19#2) #3}
\newcommand{\pl}[3]{{\sl Phys. Lett.} {\bf #1} (19#2) #3}
\newcommand{\pr}[3]{{\sl Phys. Rev.} {\bf #1} (19#2) #3}
\newcommand{\prl}[3]{{\sl Phys. Rev. Lett.} {\bf #1} (19#2) #3}
\newcommand{\zp}[3]{{\sl Z. Phys.} {\bf #1} (19#2) #3}
\newcommand{\ijmp}[3]{{\sl Int. J. Mod. Phys.} {\bf #1} (19#2) #3}
\newcommand{\sjnp}[3]{{\sl Sov. J. Nucl. Phys.} {\bf #1} (19#2) #3}
\newcommand{\jop}[3]{{\sl Journal of Physics} {\bf #1} (19#2) #3}
\newcommand{\ibid}[3]{{\sl ibid.} {\bf #1} (19#2) #3}
\newcommand{\ej}[3]{{\bf #1} (19#2) #3}
\newcommand{\hep}[1]{{\sl hep--ph/}{#1}}

\end{document}